\documentclass{article}
\usepackage{graphicx} 
\usepackage[left=1.5cm,top=1.5cm,right=1.5cm,bottom=1.5cm]{geometry}
\usepackage{amsmath}
\usepackage{amssymb}
\usepackage[title]{appendix}
\usepackage{listings}
\usepackage{indentfirst}
\usepackage[dvipsnames]{xcolor}
\usepackage{blindtext}
\usepackage{float}
\usepackage{siunitx}
\usepackage{booktabs}
\usepackage{hyperref}
\usepackage{caption,pdfpages}
\usepackage{cite} 
\hypersetup{colorlinks=true,linkcolor=blue,citecolor=blue,urlcolor=blue}

\title{A new analytical finding for the Born-Infeld integral potential}

\author{Leonardo P. G. De Assis$^2$, Priscila Valdênia dos Santos$^3$,\\ Sergio de Oliveira Vellozo$^4$, José Abdala Helayël-Neto$^1$ \\ \\
$^1$Centro Brasileiro de Pesquisas Físicas - CBPF - Rio de Janeiro - Brazil, \\ $^2$ULO-Math and Physics, Stanford University - CA - USA, \\ $^3$Universidade Federal de Pernambuco - UFPE - Caruaru - Pernambuco - Brazil, \\ $^4$Instituto Militar de Engenharia - IME - Rio de Janeiro - Brazil }

\date{\today}

\begin{document}
\maketitle

\begin{abstract}
    A new finding regarding the Born-Infeld potential for short range  distances from the charge revealed that its integral form can be almost exact written as the difference between a Coulomb-type potential and a Yukawa-type potential. This allowed for an immediate physical interpretation in terms of classical vacuum polarization and an effective mass. Subsequently, a new $\widetilde U_B(r)$ is proposed to cover the entire space. This formulation is then applied to calculate, in a closed manner, the first-order energy shifts in hydrogen-like atoms; a comparison of these shifts with high-precision spectroscopic data provides a model-dependent lower bound for the maximum field intensity parameter \(b\).
\end{abstract}

\section{Introduction}

The Born--Infeld theory of electromagnetism \cite{Born33,Born34,Greiner}, originally conceived to resolve the infinite self-energy of a point charge, represents a nonlinear extension of Maxwell's equations \cite{Panza,
Dehghani}. Its Lagrangian, in the weak-field limit, contains quartic field operators which are also characteristic of the one-loop Euler--Heisenberg effective action in QED \cite{Berestetskii,Akhiezer,Kramida,Hänsch}, suggesting a connection between classical nonlinearity and quantum vacuum effects. The central difficulty in applying the theory to physical problems, such as atomic spectra, is that the exact electrostatic potential for a point charge is not available in closed form \cite{Vellozo1,Vellozo2,AlBadawi2025}. In this paper we shall hence develop a practical method based on a simple analytic function $\widetilde U_B$ for the potential, one which correctly reproduces the required asymptotic behavior (Coulombic at large distances, finite at the origin) and allows for a straightforward calculation of perturbative atomic energy shifts, from which we can derive experimental constraints on the fundamental BI parameter \(b\).

\subsection*{Conventions and units}
We use SI units throughout this work (so that \(\varepsilon_0\) appears explicitly). In sections where a comparison with the Euler--Heisenberg effective Lagrangian is helpful, we momentarily display expressions in natural units (\(\hbar=c=1\)) and indicate the conversion back to SI.

\section{BI Potential for r near origin }

The exact BI potential of a point like charge $e$ at $r=0$ is writen as the electric field space integral with $q=\frac{e}{4 \pi \varepsilon_0}$.

\begin{equation}
U_B(r) = q \int^{\infty}_r \frac{du}{\sqrt{r_b^4+u^4}}
\label{BI01}
\end{equation}

The two important properties for us are, as can be seen by equation (\ref{BI01}): 
\vspace{0.5cm}

for $r \gg r_b \rightarrow \frac{q}{r}$ (Coulomb like potential) 

for $r=0 \rightarrow \beta \frac{q}{r_b}$ (finite)
\vspace{0.5cm}

\noindent where $\beta$ is the dimensionless characteristic number defined as the integral below,

\begin{equation*}
    \beta = \int^{\infty}_0 \frac{du}{\sqrt{1+u^4}}=\frac{4\Gamma^2(5/4)}{\sqrt{\pi}}=1.854074677\ldots
\end{equation*}

\noindent and $r_b$ is the characteristic BI short-distance length.

The aim is to change the integral by analytical functions.  For this task we write $\int^{\infty}_r$ as $\int^{\infty}_0$ - $\int^{r}_0$. The $\int^{r}_0$ is suitable to make aproximations. Rewriting equation (\ref{BI01}) in terms of dimensionless variable $x=\frac{r}{r_b}$:

\begin{equation}
U_B(r) = \frac{q}{r_b} \left\{ \beta-\int^{x}_0 \frac{du}{\sqrt{1+u^4}}
\label{BI02} \right\}
\end{equation}

Equation (\ref{BI02}) is general and suitable to compute $U_B$. For $x \ll 1$, the integral $\int^{x}_0 \frac{du}{\sqrt{1+u^4}} \approx \int^{x}_0 du = x$ and the equation (\ref{BI02}) is almost in its final version. 

\begin{equation}
U_B(r) \approx \frac{q}{r_b} \left\{ \beta-x \right\}=\frac{q}{r_b} \left\{ \frac{\beta x-x^2 }{x} \right\}
\label{BI03}
\end{equation}

The final step is reached when, in the numerator, 1 is added and subtracted, $x^2$ is considered very small compared to 1 and $1-\beta x \approx e^{-\beta x}$. So, we have:

\begin{equation*}
 \frac{\beta x-x^2 }{x} = \frac{1-1+\beta x-x^2 }{x} \approx \frac{1}{x}-\frac{e^{-\beta x}}{x}
\label{BI04}
\end{equation*}

 Finally, the BI potential can be written as the diference of a Coulomb-like potential and a Yukawa-like potential.

\begin{equation}
 \widetilde U_B(r) \approx q\left\{ \frac{1}{r}-\frac{e^{-\frac{\beta}{r_b} r}}{r} \right\}
\label{BI05}
\end{equation}

\begin{figure}[H]
    \centering
    \includegraphics[height=70mm,width=85mm]{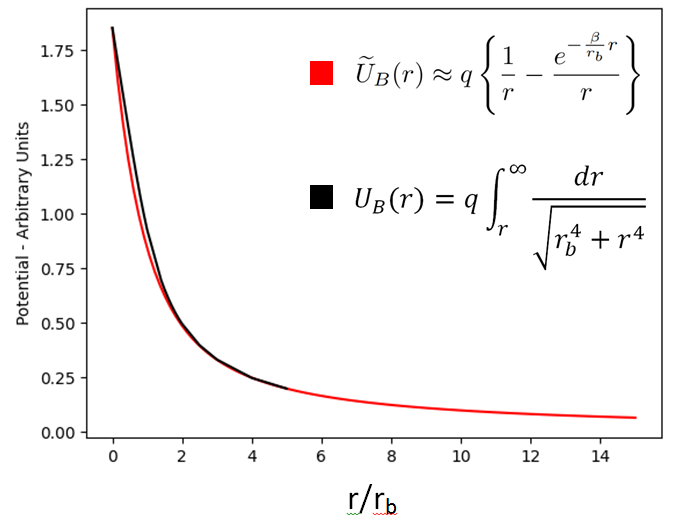}
    \caption{A comparison of the exact BI potential (computed via numerical integration) and the Coulomb-plus-Yukawa extended function given in Eq.(\ref{BI05}).}
    \label{fig:G1}
\end{figure}

Although it was derived under restrictions ($r \ll r_b$), we can see that it can be much more general. It smoothly recovers the Coulomb potential for $r \gg r_b$. Even the transition region $r \approx r_b$, proved to be very close to the BI potential, as can be seen in Figure \ref{fig:G1}. 

Note that the departure of the Yukawa-type potential from the Coulomb-type potential is most pronounced at large distances from the source charge, not at small ones. Our representative potential $\widetilde{U}_B$ brings together both sectors, and we need to interpret it from both a classical and a quantum perspective. This is also expressed by the statement that when an interaction is mediated by the exchange of a massive gauge particle the interaction must have a finite range that gets smaller with increasing mass. Consequently, since photons are expected to cross billions of light years of deep space, one can get an upper bound on the photon mass from cosmological considerations. As we know in other contexts, the introduction of a mass term usually breaks the U(1) gauge symmetry of the field theory, unless it appears as a result of the Higgs mechanism. 

Therefore, we choose the function from equation (\ref{BI05}), which emerged from the original BI integral potential, as representative of it as well all its consequences. The intractable original BI integral potential will be replaced by a compact Coulomb-plus-Yukawa potential with \(r_b\) being the characteristic BI short-distance length and \(\beta\) a dimensionless constant which we must choose to match the small-\(r\) behavior of the true BI integral.

\subsection{Electric field at the origin}
 
Differentiating the potential in equation (\ref{BI05}) gives the electric field. Evaluating the expression \(E(r)=-d\widetilde U_B/dr\) in the limit \(r\to0\) yields a finite result:

\[
\begin{split}
E(0)=\lim_{r \to 0} -\frac{d\widetilde U_B(r)}{dr} = q\lim_{r \to 0} \frac{1-(1+\frac{\beta r}{r_b})e^{-\frac{\beta r}{r_b}}}{r^2}= \\ \approx q\lim_{r \to 0} \frac{1-(1+\frac{\beta r}{r_b})(1-\frac{\beta r}{r_b})}{r^2}= \beta^2\frac{q}{r_b^2}
\end{split}
\]

Identifying the BI maximum field strength \(b\) with this finite field at the origin gives the central mapping employed throughout this manuscript:

\begin{equation}
E(0)=\beta^2\frac{e}{4\pi\varepsilon_0\,r_b^2}=b.
\label{eq:E0}
\end{equation}

This expressions follows directly from the equation (\ref{BI05}) and must be used consistently. They differ by a constant from original BI. Our interest lies in the potential because in quantum mechanics that's what's important.

\section{An interlude: classical polarization interpretation}

A more direct physical approach is to interpret the structure of the $\widetilde U_B$ potential in equation (\ref{BI05}). It is the sum of two parts,

\[
\widetilde{U}_B = \underbrace{\frac{e}{4\pi\varepsilon_0}\frac{1}{r}}_{\text{Coulomb}} - \underbrace{\frac{e}{4\pi\varepsilon_0}\frac{e^{-\beta r/r_b}}{r}}_{\text{Screening}},
\]

The second term can be pictured as arising from a short-range screening cloud of opposite charge density. This provides a classical analogue of quantum vacuum polarization. The analogy is, of course, not exact; the underlying mechanism here is a classical field nonlinearity rather than the creation and annihilation of virtual particle-antiparticle pairs, and as one can see in the next Section, the operator structure differs numerically from that of one-loop QED.

Polarized charge of opposite signal is distributed radially, and this distribution can be obtained from:

\[
\nabla^2 \left\{q \frac{e^{-\beta\frac{r}{r_b}}}{r} \right\} = - \frac{\rho}{\epsilon_o}  \Longrightarrow  dQ=-e x e^{-x} dx 
\]

\begin{figure}[H]
    \centering
    \includegraphics[height=70mm,width=85mm]{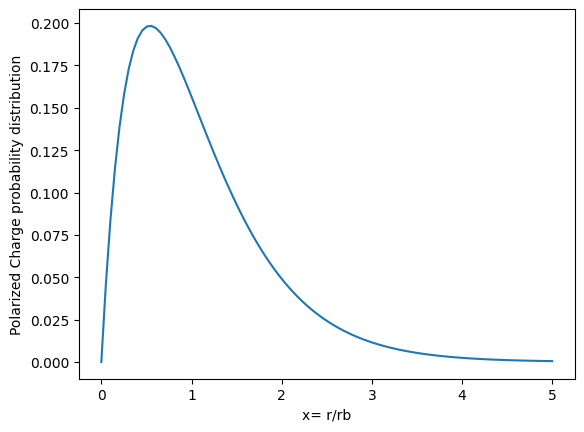}
    \caption{Radial distribution of polarized charge.}
    \label{fig:G2}
\end{figure}

The total charge intensity is, by integration $Q=-e$. The distribution function is $p(x)=xe^{-x}$ and its maximum accurs at $x=1$, this is, for $r_b/\beta$. This means that the polarization is distributed very close to the point charge and quickly goes to zero after $r_b$ \cite{DeBenedictis2025}. The figure \ref{fig:G2} shows the polarized charge distribution as funtion of $x=\beta r/r_b$.

Furthermore, it is evident that the $\beta$ factor arises from the short-range sector (Yukawa-type potential), which also vanishes as $r_b$ tends to zero (point like charge). This preserves the pure Coulomb potential representing the massless photon. Thus, we are tempted to suggest that $\beta$ is a short-range sector signature absent in the pure Coulomb potential.

We discussed and concluded that, from the classical point of view, $\widetilde{U}_B$ explicitly contains a term that we interpret as a classical polarization of the vacuum.

In the next section, we will discuss whether the linearization of BI Lagrangian, still within a purely classical view, would be capable of generating mass.

\section{Quantitative Analysis of the Classical and Quantum Analogy}

This section contains two parts: (i) the linearization of the BI Lagrangian in field-theory terms, (ii) the effective field theory (EFT) weak-field comparison to Euler--Heisenberg QED. We leave the perturbative calculations for spectroscopy which link our $\widetilde{U}_B$ to experimental constraints \cite{Goulart2026} to the section 5.

\subsection{Linearization of the BI Lagrangian}
\label{subsec:linearization}

The BI Lagrangian, expressed in terms of the Lorentz invariants \(F=\tfrac{1}{2}(E^2-c^2 B^2)\) and \(G=c\,\mathbf{E}\cdot\mathbf{B}\) (taking \(c=1\) in this section for simplicity), reads
\begin{equation}
\mathcal{L}_{\mathrm{BI}}(F,G)=b^2\left(1-\sqrt{1-\frac{2F}{b^2}-\frac{G^2}{b^4}}\right).
\label{eq:LBI}
\end{equation}

Let us consider small fluctuations \(\boldsymbol{e},\boldsymbol{b}\) propagating on top of a constant, homogeneous background field \((\mathbf{E}_0,\mathbf{B}_0)\), such that \(\mathbf{E}=\mathbf{E}_0+\boldsymbol{e}\) and \(\mathbf{B}=\mathbf{B}_0+\boldsymbol{b}\). Expanding the invariants up to second order in the perturbations,
\begin{align}
\delta F &= \mathbf{E}_0\cdot\boldsymbol{e} - \mathbf{B}_0\cdot\boldsymbol{b} + \tfrac12(\boldsymbol{e}^2-\boldsymbol{b}^2), \\
\delta G &= \mathbf{E}_0\cdot\boldsymbol{b} + \mathbf{B}_0\cdot\boldsymbol{e} + \boldsymbol{e}\cdot\boldsymbol{b},
\end{align}
a systematic Taylor expansion of the Lagrangian density up to quadratic order in the fluctuations yields

\begin{equation}
\begin{array}{rll}
 \mathcal{L}^{(2)} & = & \mathcal{L}_G (\boldsymbol{e}\cdot\boldsymbol{b}) +\tfrac{1}{2}\mathcal{L}_F (\boldsymbol{e}^2 - \boldsymbol{b}^2) + \\
  &  & + \tfrac{1}{2}\mathcal{L}_{FF}(\mathbf{E}_0\cdot\boldsymbol{e} - \mathbf{B}_0\cdot\boldsymbol{b})^2+ \\
  &  & +\tfrac{1}{2}\mathcal{L}_{GG}(\mathbf{E}_0\cdot\boldsymbol{b} + \mathbf{B}_0\cdot\boldsymbol{e})^2 \\
  &  & + \mathcal{L}_{FG}(\mathbf{E}_0\cdot\boldsymbol{e} - \mathbf{B}_0\cdot\boldsymbol{b})(\mathbf{E}_0\cdot\boldsymbol{b} + \mathbf{B}_0\cdot\boldsymbol{e}),
\end{array} 
\end{equation}

\vspace{0.5cm}

\noindent where the partial derivatives \(\mathcal{L}_F \equiv \partial\mathcal{L}/\partial F\), \(\mathcal{L}_{FF} \equiv \partial^2\mathcal{L}/\partial F^2\), etc., are evaluated at the unperturbed background configuration \((F_0, G_0)\).

Specializing to a purely magnetic background (\(\mathbf{E}_0 = 0\), \(\mathbf{B}_0 \neq 0\)), one has \(F_0 = -B_0^2/2\) and \(G_0 = 0\). Due to the parity symmetry of the theory, \(\mathcal{L}_G\big|_0 = \mathcal{L}_{FG}\big|_0 = 0\), and the fluctuation Lagrangian simplifies to
\begin{equation}
\mathcal{L}^{(2)} = \tfrac12 c_1(\boldsymbol{e}^2-\boldsymbol{b}^2) - d_1(\mathbf{B}_0\cdot\boldsymbol{b})^2 - d_2(\mathbf{B}_0\cdot\boldsymbol{e})^2,
\label{eq:L2}
\end{equation}

\vspace{0.5cm}

\noindent where the coefficients are given directly by the background derivatives:
\begin{align}
c_1 &= \mathcal{L}_F\Big|_0 = \left(1+\frac{B_0^2}{b^2}\right)^{-1/2}, \label{eq:c1} \\
d_1 &= -\frac{1}{2}\mathcal{L}_{FF}\Big|_0 = -\frac{1}{2b^2}\left(1+\frac{B_0^2}{b^2}\right)^{-3/2}, \label{eq:d1} \\
d_2 &= -\frac{1}{2}\mathcal{L}_{GG}\Big|_0 = \frac{1}{2b^2}\left(1+\frac{B_0^2}{b^2}\right)^{-1/2}. \label{eq:d2}
\end{align}

From the modified constitutive relations \(\boldsymbol{d} = \partial\mathcal{L}^{(2)}/\partial\boldsymbol{e}\) and \(\boldsymbol{h} = -\partial\mathcal{L}^{(2)}/\partial\boldsymbol{b}\), Maxwell's equations for plane-wave fluctuations \(\propto e^{i(\mathbf{k}\cdot\mathbf{r}-\omega t)}\) lead to the characteristic equation governing the propagation modes \cite{Nenad,Hedvika2021}. Because Born--Infeld electrodynamics satisfies the Boillat--Pleba\'nski conditions for exceptional waves (i.e., complete absence of birefringence), both independent photon polarization states share the exact same dispersion relation in this magnetic background \cite{AraujoFilho2026,AraujoFilho2025,Sokolov2025}:
\begin{equation}
\omega^2 = k^2\left[1-\frac{(\mathbf{B}_0\times\mathbf{k})^2}{k^2(b^2+B_0^2)}\right] = \frac{k^2}{n(\theta)^2},
\end{equation}
with the direction-dependent refractive index given by
\begin{equation}
n(\theta)^2 = \frac{b^2+B_0^2}{b^2+B_0^2\cos^2\theta},
\end{equation}
where \(\theta\) is the angle between the propagation vector \(\mathbf{k}\) and the background magnetic field \(\mathbf{B}_0\).

\subsection{EFT comparison: weak-field expansions and coefficient matching}
\label{subsec:EFT}

A weak-field expansion of the BI Lagrangian yields, to quartic order,
\[
\mathcal{L}_{\mathrm{BI}}\approx \tfrac12(E^2-B^2) + \frac{1}{8b^2}(E^2-B^2)^2 + \frac{1}{2b^2}(\mathbf{E}\cdot\mathbf{B})^2 + \mathcal{O}(b^{-4}).
\]
For comparison, the Euler--Heisenberg one-loop effective Lagrangian of QED is
\[
\mathcal{L}_{\mathrm{EH}}\approx \tfrac12(E^2-B^2) + \frac{2\alpha_{\mathrm{em}}^2}{45m_e^4}\big[(E^2-B^2)^2 + 7(\mathbf{E}\cdot\mathbf{B})^2\big] + \mathcal{O}(m_e^{-8}).
\]
Matching the coefficient of the \((E^2-B^2)^2\) term formally gives a relation for \(b\),
\[
\frac{1}{8b^2}=\frac{2\alpha_{\mathrm{em}}^2}{45m_e^4}\quad\Rightarrow\quad b_{\mathrm{EFT}}=m_e^2\frac{\sqrt{45}}{4\alpha_{\mathrm{em}}}\sim 9\times10^{19}\ \mathrm{V/m,}
\]
after conversion to SI units. It must be highlighted that this is a heuristic scale-matching; the mismatch in the operator-ratios (BI is \(1{:}4\), whereas EH is \(1{:}7\)) means the theories are not equivalent, and consequently predictions for observables sensitive to the $(\mathbf{E}\cdot\mathbf{B})^2$ operator will necessarily differ.

Once this new representation of BI potential is established, it is interesting to apply it to systems where there is ample experimental data \cite{Ahumada2026}. For the next section, we chose the hydrogen atom to be used as a laboratory test. It is a simple structure that allows for exact solutions. Within the context developed in the previous sections, we establish limits for the maximum intensity of the $b$ field.  

\section{The hydrogen atom as a laboratory and the static new Potential}

We now have all the tools required to compute the effect of the BI nonlinearity on atomic energy levels. The deviation of our $\widetilde U_B$ potential from the pure Coulomb potential, \(\Delta\widetilde U(r)=q^2e^{-\beta r/r_b}/r\), is treated as a time-independent perturbation to the hydrogen atom Hamiltonian \cite{Porto2026,Cheng2025}. For the ground state, the first-order energy shift \(\delta\mathcal{E}_1\) is then given by the expectation value,

\[
\delta\mathcal{E}_1=\langle\phi_{1s}|\frac{q^2e^{-\beta r/r_b}}{r}|\phi_{1s}\rangle.
\label{delta E}
\]

The details of the integral are straightforward (see Appendix \ref{app:integral}), and the final result, after eliminating the length scale \(r_b\) in favor of the maximum field \(b\), shows that the energy shifts for s- and p-states have a characteristically different dependence on \(b\) (namely, \(\delta\mathcal E_{1s}\propto 1/b\) while \(\delta\mathcal E_{2p}\propto 1/b^2\)). By requiring that this calculated shift be smaller than the current experimental uncertainty in hydrogen spectroscopy, we arrive at a lower bound on the BI field strength,

\[
b_{\mathrm{spec}}\gtrsim(1.23\text{--}1.36)\times10^{24}\ \mathrm{V/m}.
\]

It must be highlighted, however, that this result is model-dependent \cite{Chen2026}; a definitive calculation would require a full treatment including Dirac wavefunctions, the Lamb shift, and finite-nuclear-size effects.

The hydrogen atom admits several levels of approximation. A Bohr-model estimate is pedagogically useful but of limited quantitative value. The Schrödinger-level perturbative calculation (which we have performed here) is the minimal quantum test for any $\widetilde U_B$-induced shifts. The final, definitive calculation, however, should employ Dirac wavefunctions and include Lamb, recoil, and finite-proton-size corrections.

\subsection{Bohr-BI atom model (pedagogical)}

A short account; Bohr's complete derivation is presented in Appendix~\ref{app:bohr}. Under the $\widetilde U_B$ potential, the Bohr energy gains a small correction that scales like \(\sim r_b/a\); this is not a reliable estimate for precision spectroscopy bounds \cite{Leite,Akmansoy,Mazharimousavi,Kiessling,Franklin}.

\subsection{The Schrödinger ground state and first-order perturbation}

The unperturbed ground-state wavefunction \cite{Leite} is
\[
\phi_{1s}=\frac{e^{-r/a}}{\sqrt{\pi a^3}},\qquad a=\frac{\hbar^2}{m q^2},
\]
and a first-order perturbation with the potential \(\Delta\widetilde U\) leads directly to the integral evaluated in Appendix~\ref{app:integral}.

\section{The energy level 2 (wavefunctions)}

The $2p$ sublevel wavefunctions are given by their standard textbook forms. We restore the three wavefunction expressions as tables, as in the original manuscript, for clarity.

\begin{table}[!h]
\centering
\begin{tabular}{ c|c|c|l } 
\hline
n & l & m & $\phi_{nlm} $ \\ 
\hline
  &  & 0 & $ \frac{1}{4} \left( \frac{1}{\sqrt{2\pi a^3}} \right)\left(\frac{r}{a} \right) \cos\theta e^{-r/(2a)}$ \\
 2 & 1 & +1 & $ \frac{1}{8} \left( \frac{1}{\sqrt{\pi a^3}} \right) \sin\theta \left(\frac{r}{a} \right) e^{-r/(2a)} e^{i\varphi} $ \\
  &  & -1 & $ \frac{1}{8} \left( \frac{1}{\sqrt{\pi a^3}} \right) \sin\theta \left(\frac{r}{a} \right) e^{-r/(2a)} e^{-i\varphi}$ \\
\hline
\end{tabular}
\caption{ Wave functions for sublevel 2p.}
\end{table}

\section{Hydrogen experimental data and theoretical predictions}

We compare our analytic shifts to precise hydrogen data (from NIST and Hänsch). The small perturbation \(\delta\mathcal E\) must be smaller than the experimental error to avoid conflict. The simple survey below reproduces the original table form.

\begin{table}[h]
  \centering
\begin{center}
\begin{tabular}{c|c|c|c} 
 \hline
   Transition & $\mathcal{E}_{n}$ & $\delta \mathcal{E}_{n}$ & $b >$  \\ 
              &  (eV)             &  (eV)                    & (V/m) \\
  \hline
              & 13.598434599702 & $1.2 \times 10^{-11}$ & $1.36 \times 10^{24}$ \\ 
  \hline
  $2p \rightarrow 1s$ & 10.21272576272 & $9.9 \times 10^{-12}$ & $1.23 \times 10^{24}$ \\ 
  \hline
\end{tabular}
\end{center}
\caption{Estimating the low bound for $b$.}
 \label{tab:Transition}
\end{table}

These numbers are of the same order as found previously; again, they are model-dependent bounds.

\section{Conclusion}
\label{sec:conclusion}

The principal contribution of this paper has been to show that the analytically tractable Coulomb-plus-Yukawa potentials, $\widetilde U_B$, emerge from original untractable BI potential. It captures the two essential electrostatic features of Born--Infeld theory: the long-range Coulombic tail and the regularization of the field at the origin. As an immediate consequence, the classical polarization of vacuum appears explicitly represented by the Yukawa sector of the $\widetilde U_B$ potential. From this starting point, we have derived closed-form, first-order perturbative expressions for hydrogenic energy shifts. The utility of these formulae is that they make the dependence of atomic observables on the BI scale \(b\) entirely explicit (for example, \(\delta E_{1s}\propto 1/b\) and \(\delta E_{2p}\propto 1/b^2\) in the regime where \(r_b\) is small), and thus provide a simple, reproducible method for translating high-precision spectroscopic data into model-dependent constraints on the fundamental parameters of the theory. 

Complementing this static, phenomenological analysis, we have also documented the field-theoretic linearization of the BI Lagrangian and an EFT-style weak-field comparison with the Euler--Heisenberg effective action of QED. These two theoretical strands serve different but complementary roles. The linearization clarifies why the classical BI vacuum behaves like a field-dependent dielectric and why BI does not induce a photon mass gap, while the quartic weak-field expansion explains, at the level of operator structure, why BI and one-loop QED produce qualitatively similar nonlinear optical effects. Making both calculations explicit in the same manuscript helps to identify precisely which aspects of QED the classical BI model can mimic and where it necessarily differs (for example, via the mismatch in relative operator coefficients and the presence of QED nonlocal vacuum-polarization pieces).

A noteworthy result of this combined analysis is a quantitative tension between two operational uses of the parameter \(b\): the EFT-benchmark value \(b_{\mathrm{EFT}}\) obtained by matching quartic coefficients, and the spectroscopy-derived, $\widetilde U_B$-dependent lower bound \(b_{\mathrm{spec}}\) obtained from hydrogen-level shifts. This tension is informative rather than paradoxical. It shows that equating the EFT matching scale with the literal short-distance BI scale in a simple Coulomb-plus-Yukawa model is not guaranteed, and that resolving that ambiguity requires either (i) validating the $\widetilde U_B$ against the exact BI potential in the radius range sampled by atomic wavefunctions or (ii) performing Dirac-level, radiative- and finite-nuclear-size-corrected spectral calculations. The manuscript hence supplies both a concise diagnostic (the analytic expressions) and a clear prescription for the numerical and experimental work needed to convert an $\widetilde U_B$-dependent constraint into a definitive limit.

The work has several practical benefits. First, the $\widetilde U_B$ and the closed-form expressions provide a quick, reproducible way for researchers to estimate the size of BI-induced atomic effects and to perform sensitivity studies across different systems (e.g., muonic atoms, high-\(Z\) hydrogenic ions). Second, the explicit linearization and EFT discussion offer a transparent bridge between classical nonlinear-field intuition and quantum vacuum phenomenology, which can guide where classical simulations might be a good proxy for quantum calculations (and where they will fail). Third, the paper collects numeric benchmarks and a sensitivity table that quantify how modest changes in short-distance modelling alter inferred bounds, information that is essential for an honest interpretation of any constraint derived from spectroscopy.

At the same time, the manuscript carefully states its limitations. The numerics and bounds reported are explicitly $\widetilde U_B$-dependent and based on nonrelativistic first-order perturbation theory; they should hence be regarded as informative, model-specific constraints rather than definitive exclusions of Born--Infeld electrodynamics. We have highlighted the minimal, high-leverage follow-up calculations required to firm up the conclusions (numerical evaluation of the exact BI potential, Dirac-level perturbation theory, inclusion of Lamb and proton-size effects, and targeted studies of systems whose wavefunctions sample shorter radii). When those steps are performed, the analytic toolkit developed here will make it straightforward to update and sharpen the resulting limits.

In summary, the manuscript contributes a useful, transparent analytic framework and an explicit set of diagnostics that together clarify how classical BI nonlinearities relate to quantum vacuum effects and how those relations can be probed with precision atomic spectroscopy. The results are reproducible and appropriately cautious: they add value to the literature by supplying both practical formulae for phenomenologists and a clear program for the numerical and experimental work needed to make the model-dependent bounds decisive.

\section*{Notice}
This article did not use artificial intelligence tools.

\begin{appendices}

\section{The integral for ground state energy uncertainty}
\label{app:integral}

Redefining $q^2$ as   $e^2 / 4\pi \epsilon_0$, the ground-state perturbation integral is
\[
\delta \mathcal{E}_1 = \langle \phi_1|\frac{q^2 e^{-\beta r/r_b}}{r}|\phi_1\rangle
=4\pi q^2\int_0^\infty \frac{e^{-2r/a}}{\pi a^3} \, r \, e^{-\beta r/r_b} dr.
\]
Let us define \(\gamma\equiv 2/a+\beta/r_b\). Then the integral becomes
\[
\delta \mathcal{E}_1 = \frac{4q^2}{a^3}\int_0^\infty r e^{-\gamma r} dr=\frac{4q^2}{a^3}\frac{1}{\gamma^2}.
\]
For the case of interest, \(r_b\ll a\), we can take \(\gamma\approx\beta/r_b\) and obtain
\[
\delta\mathcal{E}_1\approx\frac{4q^2}{a^3}\frac{1}{b}.
\]
Eliminating \(r_b^2\) via equation \eqref{eq:E0} gives the \(1/b\) scaling and the algebraic prefactor quoted in the main text; a fully explicit numeric evaluation is straightforward and reproducible by substituting the SI constants.

\section{A particle associated with the Yukawa-type potential? (speculative)}
\label{app:particle}

A heuristic identification of the Yukawa inverse length $\frac{\beta}{r_b}$ with a Compton mass \(m_0\) gives

\[
\frac{m_0c}{\hbar}=\frac{\beta}{r_b}\quad\Longrightarrow\quad m_0=\frac{\hbar}{c}\sqrt{\frac{q^2}{e}}\sqrt{b}.
\]

By means of equation \eqref{eq:E0} this can be expressed in terms of \(b\); the resulting masses are enormous (many times \(m_e\)) and this identification is speculative. We keep it as a parametrization only.

\begin{table}[h]
\begin{center}
\begin{tabular}{c|c}
 \hline
   $b$ (V/m) & $m_0/m_e$  \\ 
  \hline
  $10^{20}$ & 188 \\ 
  $10^{22}$ & 1886 \\ 
  $10^{24}$ & 18867 \\ 
  \hline
\end{tabular}
 \caption{Heuristic masses associated to the Yukawa inverse length $\frac{\beta}{r_b}$.}
\label{TB2}
\end{center}
\end{table}

\section{Bohr-model details}
\label{app:bohr}

We consider that the hydrogen atom, in particular its ground state energy, can be a suitable natural system to test theories and provide the parameters of interest. In the more specific case, we intend to study how hydrogen the e-p potential, finite and derived from a nonlinear theory, can affect the standard coulombian value of this state.

\subsection{The Schr\"odinger wave equation for hydrogen}

 The Schr\"odinger wave equation for hydrogen on ground state energy level is written as:

\begin{equation}
-\alpha \nabla^{2} \phi_1 + U_C \phi_1 = \mathcal{E}_1 \phi_1
\label{A1}
\end{equation}

Where $\phi_1$ is the ground state wave function:

\begin{equation}
    \phi_1 = \frac{e^{-\frac{r}{a}}}{\sqrt{\pi a^3}}
\label{A2}
\end{equation}

$U_C$ is the Coulomb electical potential energy:

\begin{equation}
    U_C = -\frac{q^2}{r}
\label{A3}
\end{equation}

$\mathcal{E}_1$ is the ground state energy:

\begin{equation}
    \mathcal{E}_1 = -\frac{mq^4}{2\hbar^{2}}
\label{A4}
\end{equation}

\noindent $\alpha = \hbar^2/2m$, $a=\frac{\hbar^2}{mq^2}$.
\vspace{0.5cm}

All this is well known and needs no comment.

\subsection{The Born-Infeld hydrogen ground state}

\indent In the nonlinear BIE framework, the potential energy between proton and electron is no longer coulombian and is finite at origen. It looks like this:

\begin{equation}
    U_B(r) = -q^2\int_{r}^{\infty} \frac{dr}{\sqrt{r^4+r_b^4}}   
\label{A5}
\end{equation}

The Schr\"odinger wave equation, with BI potential energy, is similar to equation \ref{A1}:

\begin{equation}
-\alpha \nabla^{2} \phi + U_{B} \phi = \mathcal{E} \phi
\label{A6}
\end{equation}

\noindent where $\mathcal{E}$ and $\phi$ are the unknow ground state energy  and  wave function, respectively. 

Considering that the difference between equations (\ref{A1}) and (\ref{A5}) resides, mainly, in the non-Coulomb form of the potential energy and that this difference happens very close to the electric charge, we can rise the assumption that $\phi$ is little disturbed and we take $\phi \approx \phi_0$ as first approximation.

\subsection{The shift prediction for BI ground state potential}

The new function $\widetilde{U}_B$ is the sum of two terms: one essentially Coulombian, $- \frac{q^2}{r}$ and the other similar to a potential like Yukawa  $\frac{q^2}{r} e^{-\beta\frac{r}{r_b}}$. Inserting $\widetilde{U}_B$ in Schr\"odinger wave equation:

\begin{equation}
-\alpha \nabla^{2} \phi + \widetilde{U}_B \phi = \mathcal{E} \phi
\label{A12}
\end{equation}

Once again, taking $\phi \approx \phi_1$:

\begin{equation}
-\alpha \nabla^{2} \phi_1 -\frac{q^2}{r} (1-e^{-\beta \frac{r}{r_b}}) \phi_1 = \{ \mathcal{E}_1+\delta \mathcal{E}_1 \} \phi_1
\label{A13}
\end{equation}

Multiplying on the left side by $\phi_1$, integrating over the whole volume, we arrive at:

\begin{equation}
\langle \phi_1| -\alpha \nabla^{2} -\frac{q^2}{r} -\mathcal{E}_1| \phi_1\rangle =\langle \phi_1 |-\frac{  q^2e^{ -\beta \frac{r}{r_b} }  }{r} + \delta \mathcal{E}_1 |\phi_1 \rangle
\label{A14}
\end{equation}

The left side is zero. The right side must also be zero and considering that $\phi_1$ is normalized, $\langle \phi_1 |\phi_1\rangle = 1$, we write:

\begin{equation}
\delta \mathcal{E}_1=\frac{\langle \phi_1 |\frac{  q^2e^{-\beta \frac{r}{r_b}}}{r} |\phi_1\rangle}{\langle \phi_1 |\phi_1\rangle}=\langle \phi_1 |\frac{  q^2e^{-\beta \frac{r}{r_b}}}{r} |\phi_1\rangle
\label{A15}
\end{equation}

The integral of equation (\ref{A15}) is not difficult to calculate analytically. Considering $r_b \ll a$, the shift can be written in terms of fundamental constants of physics $e, \varepsilon_0, \hbar, m$ and the BI fundamental parameter field $b$. Equation (\ref{A15}) can be more properly written in terms of fundamental constants and the main parameter of BI electrodynamics, the maximum field strength $\textit{b}$. This field $\textit{b}$ is related to $r_b$ by $b=\beta^2 \frac{e}{4\pi\epsilon_0 r^2_b}$.

\begin{equation}
\langle \phi_1 |\frac{  q^2e^{-\beta \frac{r}{r_b}}}{r} |\phi_1\rangle = \frac{4q^4}{e a^3 } \frac{1}{b}  \hspace{3em}    for \hspace{2em} a\gg r_b
\label{A16}
\end{equation}

By equation (\ref{A16}) it is easy to see that $\delta \mathcal{E}_1 \propto \frac{1}{b}$. This is a significant result since in Maxwell electrodynamics the field strength $b \to \infty$ and $\delta \mathcal{E}_1 \to 0$.

Finally, we can write $\delta \mathcal{E}_1$ more simply   :

\begin{equation}
\delta \mathcal{E}_1 = \frac{8}{\beta^2}\frac{e}{4\pi\epsilon_0 a^2}\mathcal{E}_1\frac{1}{b}\approx \frac{1.2 \times 10^{12}}{b} \mathcal{E}_1
\label{A17}
\end{equation}

\noindent with $b$ in MKS units, $\mathcal{E}_1$ and $\delta \mathcal{E}_1$ in eV.

We can regard equation (\ref{A17}) as a small contribution to ground state energy.

Rewriting $\mathcal{E}$:

\begin{equation}
\mathcal{E} \approx \{ 1 + \frac{8}{\beta^2}\frac{e}{4\pi\epsilon_0 a^2}\frac{1}{b} \} \mathcal{E}_1
\label{A18}
\end{equation}

\end{appendices}


\begin{thebibliography}{99}
\bibitem{Born33}
M. Born, Foundations of the New Field Theory, Nature 132 (1933) 282.
\bibitem{Born34}
M. Born, L. Infeld, Proc. R. Soc. London Ser. A 144 (852) (1934) 425.
\bibitem{Leite}
J. Leite Lopes, \textit{A Estrutura Quântica da Matéria}, 2nd ed., 1993.
\bibitem{Akmansoy}
P. N. Akmansoy, L. G. Medeiros, Constraining Born–Infeld-like nonlinear electrodynamics using hydrogen’s ionization energy, Eur. Phys. J. C (2018) 78:143.
\bibitem{Mazharimousavi}
S. H. Mazharimousavi, M. Halilsoy, Ground state H-atom in Born--Infeld theory, Found. Phys. 42 (2012) 524–530.
\bibitem{Kiessling}
M. K.-H. Kiessling, Nonperturbative Calculation of Born–Infeld Effects on the Schrödinger Spectrum of the Hydrogen Atom, Phys. Rev. Lett. 96, 030402 (2006).
\bibitem{Franklin}
J. Franklin, T. Garon, Approximate Born–Infeld effects on the relativistic hydrogen spectrum, Phys. Lett. A 375 (2011) 1391–1395.
\bibitem{Panza}
N. Panza, Y. M. P. Gomes, J. A. Helayël, arXiv:1609.02445 (2016).
\bibitem{Dehghani}
A. Dehghani, M. R. Setare, S. Zarepour, Eur. Phys. J. Plus 137 (2022) 1–19.
\bibitem{Greiner}
W. Greiner, \textit{Quantum Electrodynamics of Strong Fields}, Plenum Press (1983).
\bibitem{Berestetskii}
V. B. Berestetskii, E. M. Lifshitz, L. P. Pitaevskii, \textit{Quantum Electrodynamics}, Pergamon Press (1980).
\bibitem{Akhiezer}
A. I. Akhiezer, V. B. Berestetskii, \textit{Quantum Electrodynamics}, Interscience (1965).
\bibitem{Vellozo1}
SO Vellozo, JA Helayël Neto, AW Smith, LPG De Assis, Standstill electric charge generates magnetostatic field under Born-Infeld electrodynamics, Int. J. Theor. Phys. 47 (2008) 2934.
\bibitem{Vellozo2}
SO Vellozo, JA Helayël Neto, AW Smith, LPG De Assis, Self-interacting electromagnetic fields and a classical discussion on the stability of the electric charge, Int. J. Theor. Phys. 48 (2009) 1905.
\bibitem{Kramida}
A. Kramida et al., NIST ASD (ver. 5.10), 2022.
\bibitem{Hänsch}
T. W. Hänsch et al., Science (2017).
\bibitem{Goulart2026}
E. Goulart and E. Bittencourt,
Metric-Induced Principal Symbols in Nonlinear Electrodynamics,
\emph{Class. Quantum Grav.} \textbf{43}, 03LT01 (2026).
doi {10.1088/1361-6382/ae415b}
\url{https://arxiv.org/abs/2508.09859}
\bibitem{Ahumada2026}
G.Ahumada Acuna, C. Bejarano, and E. Magliaro, Born-Infeld Electrogravity and Dyonic Black Holes,
\emph{Class. Quantum Grav.} \textbf{43}, 155020 (2026).
doi 10.1088/1361-6382/ae9114
\url{https://arxiv.org/abs/2511.05273}
\bibitem{Porto2026}
G.A.Porto, J.P. Pereira, E. Bittencourt, and E. Guzm\'an-Herrera, Nonlinear electrodynamics in magnetars: systematic effects on radius constraints and timing analysis,
arXiv:2605.17048 [astro-ph.HE] (2026).
\url{https://arxiv.org/abs/2605.17048}
\bibitem{AraujoFilho2026}
A. Araújo Filho, et al, Regular Black Holes in General Relativity from Nonlinear Electrodynamics with de Sitter Cores, \emph{Physics of the Dark Universe} \textbf{53}, 102396 (2026). doi{10.1016/j.dark.2026.102396}
\url{https://arxiv.org/abs/2604.20066}
\bibitem{AraujoFilho2025}
A. A. Araújo Filho, et al, Gravitational signatures of a nonlinear electrodynamics in $f(R,T)$ gravity, \emph{JCAP} \textbf{09}, 015 (2025). doi{10.1088/1475-7516/2025/09/015}
\url{https://arxiv.org/abs/2505.08718}
\bibitem{Chen2026}
C. Y. Chen, et al, Vector Horndeski black holes in nonlinear electrodynamics, \emph{Phys. Rev. D} \textbf{113}, 024027 (2026). doi{10.1103/fjqh-7gb2}
\url{https://arxiv.org/abs/2509.23134}
\bibitem{DeBenedictis2025}
A. DeBenedictis and S. Seahra, An analysis of Born-Infeld boson compact objects, \emph{Phys. Rev. D} \textbf{112}, 044003 (2025). {doi.org/10.1103/6mtm-bgy1}
\url{https://doi:10.48550/arXiv.2505.07134}
\bibitem{Sokolov2025}
V. Sokolov, Extreme black points in Born-Infeld electrodynamics, arXiv:2509.00477 [gr-qc] (2025).
\url{https://arxiv.org/abs/2509.00477}
\bibitem{Cheng2025}
L. Cheng and H. Liu, Rotating Black Holes in Einstein-Born-Infeld Theory, arXiv:2507.00879 [gr-qc] (2025).
\url{https://arxiv.org/abs/2507.00879}
\bibitem{AlBadawi2025}
A. Al-Badawi, et al., A new black hole coupled with nonlinear electrodynamics surrounded by quintessence: Thermodynamics, Geodesics, and Regge-Wheeler Potential, \emph{Chinese Journal of Physics} \textbf{94}, 102143 (2025).
doi{10.1016/j.cjph.2025.01.021}
\url{https://arxiv.org/abs/2503.00765}
\bibitem{Hedvika2021}
H. Kadlecová, Electromagnetic waves in Born--Infeld electrodynamics, Journal of Mathematical Physics (Vol.65, Issue 1), 2024 doi:10.48550/arXiv.2107.12249
\url{https://doi.org/10.1063/5.0150790}
\bibitem{Nenad}
N. Manojlovic, V. Perlick and R. Potting1, Standing wave solutions in Born-Infeld theory, \url{https://arxiv.org/abs/2006.09053v2}


\end{thebibliography}
\end{document}